\newcommand{\bea}{\begin{eqnarray}}
\newcommand{\eea}{\end{eqnarray}}
\def\rp#1#2{{#1\over#2}}
\def\bm#1{{\mbox{\boldmath$#1$\unboldmath}}}
\def\rfr#1{eq. (\ref{#1})}

\def\eqi{\begin{equation}}
\def\eqf{\end{equation}}
\def\eqia{\begin{eqnarray}}
\def\eqfa{\end{eqnarray}}
\def\lb#1{\label{#1}}
\documentstyle[ecssm]{article}
\begin{document}
\title{New proposals for the detection of the Earth's gravitomagnetic field in space-based and laboratory-based experiments}

\author{Lorenzo Iorio\dag\footnote{E-mail: Lorenzo.Iorio@ba.infn.it.}}

\affil{\dag\ Dipartimento di Fisica dell'Universit${\rm
\grave{a}}$ di Bari, Via Amendola 173, 70126, Bary, Italy}

\beginabstract
In this contribution we present two new proposals for measuring
the general relativistic gravitomagnetic component of the
gravitational field of the Earth. One proposal consists of the
measurement of the difference of the rates of the perigee $\psi$
from the analysis of the laser--ranged data of two identical
Earth'artificial satellites placed in equal orbits with
supplementary inclinations. In this way the impact of the aliasing
classical secular precessions due to the even zonal harmonics of
the geopotential would be canceled out, although the
non--gravitational perturbations, to which the perigees of
LAGEOS--type satellites are particularly sensitive, should be a
limiting factor in the obtainable accuracy. With a suitable choice
of the inclinations of the orbital planes it would be possible to
reduce the periods of such insidious perturbations so to use not
too long observational time spans. However, the use of a pair of
drag--free satellites would greatly reduce this problem, provided
that the time span of the data analysis does not excess the
lifetime of the drag--free apparatus. In the other proposal the
difference of the rotational periods of two counter-revolving
particles placed on a friction-free plane in a vacuum chamber at
the South Pole should be measured in order to extract the
relativistic gravitomagnetic signal. Among other very challenging
practical implications, the Earth's angular velocity
$\omega_{\oplus}$ should be known at a $10^{-15}$ rad s$^{-1}$
level from VLBI and the friction force of the plane should be less
than $2\times 10^{-9}$ dyne.
\endabstract
\section{The space based proposal}
\subsection{The Lense-Thirring effect}
In the weak-field and slow-motion approximation of General
Relativity, a test particle in the gravitational field of a slowly
rotating body of mass $M$ and angular momentum ${\bm J}$, assumed
to be constant, is acted upon by a non-central acceleration of the
form (Ciufolini and Wheeler, 1995; Ruggiero and Tartaglia,
2002)\eqi{\bm a}_{\rm GM}=\frac{{\bm v}}{c}\times {\bm B}_{\rm
g}\lb{eqb},\eqf in which ${\bm v}$ is the velocity of the test
particle, $c$ is the speed of light in vacuum and ${\bm B}_{\rm
g}$ is the gravitomagnetic field given by \eqi{\bm B}_{\rm
g}=\frac{2G}{c}\frac{\left[{\bm J}-3\left({\bm J}\cdot
\bm{\hat{r}}\right)\bm{\hat{r}}\right]}{r^3}\lb{gmf}.\eqf In it
$\bm{\hat{r}}$ is the unit position vector of the test particle
and $G$ is the Newtonian gravitational constant.

For a freely orbiting test particle \rfr{eqb} induces on its orbit
the so called Lense--Thirring drag of inertial frames (Ciufolini
and Wheeler, 1995). More precisely, it consists of secular
precessions of the longitude of the ascending node $\Psi$ and of
the argument of perigee $\psi$\footnote{The longitude of the
ascending node $\Psi$ is the angle, in the equatorial plane of an
inertial frame whose origin is located at the center of mass of
the central body, from a reference direction $X$, say the Aries
point $\Upsilon$, to the line of the nodes, i.e. the intersection
of the orbital plane with the equatorial plane. The argument of
perigee $\psi$ is an angle in the orbital plane from the line of
the nodes to the  direction of the perigee.} \eqia \dot\Psi_{\rm
LT} & = &
\frac{2GJ}{c^{2}a^{3}(1-e^{2})^{\frac{3}{2}}},\lb{nodolt}\\
\dot\psi_{\rm LT} & = &
-\frac{6GJ\cos{i}}{c^{2}a^{3}(1-e^{2})^{\frac{3}{2}}},\lb{perigeo}\eqfa
in which $i$, $a$ and $e$ are the inclination, the semimajor axis
and the eccentricity, respectively, of the orbit of the test body.
Such general relativistic spin--orbit effect has been
experimentally checked for the first time by analyzing the
laser--ranged data to the LAGEOS and LAGEOS II artificial
satellites in the gravitational field of the Earth with a claimed
accuracy of the order of $20\%-30\%$ (Ciufolini, 2002).
\subsection{The LAGEOS-LARES mission}
The use of the proposed LAGEOS III/LARES satellite would greatly
increase the accuracy of such space--based measurement (Ciufolini,
1986). LARES would be a LAGEOS-type satellite to be placed in the
same orbit as of LAGEOS except for the eccentricity, which should
be one order of magnitude larger, and, especially, the inclination
which should be supplementary to that of LAGEOS. Indeed, the
proposed observable would be the sum of the nodal rates
\eqi\dot\Psi_{\rm LAGEOS}+\dot\Psi_{\rm LARES}\lb{nodsum}\eqf
because, if on one hand, according to \rfr{nodolt}, the
Lense-Thirring nodal precessions are independent of the
inclination and add up in \rfr{nodsum}, on the other the aliasing
classical nodal secular precessions induced by the even zonal
harmonics of the Newtonian multipolar expansion of the terrestrial
gravitational field (Kaula, 1966), which would represent a major
source of systematic error, depend on $\cos i$ (Iorio, 2003d) and
would be canceled out in \rfr{nodsum}. However, as pointed out in
(Iorio $et\ al$, 2002a), such a perfect cancellation would not
occur due to the unavoidable orbital injection errors in the
inclination of LARES. In (Iorio $et\ al$, 2002a) a revisited
version of the LARES mission, including the orbital elements of
LAGEOS and LAGEOS II as well, is presented. It would be more
precise because it would cancel out the first four even zonal
harmonics of the geopotential irrespectively of the departures of
the LARES orbital parameters from their nominal values.
\subsection{The difference of the perigees}
The concept of satellites in identical orbits with supplementary
orbital planes could be further exploited in order to obtain a new
observable sensitive to the Lense-Thirring effect which is
complementary to and independent of the already examined sum of
the nodes. Indeed, from \rfr{perigeo} it turns out that the
gravitomagnetic shift of the perigee depends on $\cos i$ while the
classical secular precessions due to the even zonal coefficients
of the geopotential depend on $\sin i$ and $\cos^2 i $ (Iorio,
2003d). Then, for a pair of satellites in supplementary identical
orbits one could consider the difference of the perigees
\eqi\dot\psi^{(i)}-\dot\psi^{(180^{\circ}-i)}\lb{difpgi}\eqf
because, while the Lense-Thirring precessions would be equal and
opposite, the classical precessions due to the even zonal
harmonics of the geopotential would be equal and would be canceled
out in \rfr{difpgi} (Iorio, 2003a; 2003b; Iorio and Lucchesi,
2003).

With regard to the practical implementation of such an observable
it should be noticed that the proposed LAGEOS--LARES mission would
be unsuitable because the perigee is not a good observable for
LAGEOS due to the extreme smallness of the eccentricity of its
orbit ($e_{\rm LAGEOS}=0.0045$). Moreover, it should be noticed
that, contrary to the nodes, the perigees of geodetic LAGEOS--like
satellites are very sensitive to many time--dependent orbital
perturbations of gravitational and, especially, non--gravitational
origin whose periodicities are linear combinations of those of the
lunisolar ecliptical variables and of the node and the perigee of
the satellite itself. As showed in (Iorio, 2003a; Iorio and
Lucchesi, 2003), many non--gravitational perturbations are not
cancelled out by \rfr{difpgi}; if the observational time span to
be adopted for the data analysis is shorter than some of their
periods it may happen that some uncancelled time-varying
perturbations would corrupt the measurement of the investigated
Lense-Thirring effect. The optimal choice would be the use of an
entirely new pair of LAGEOS--type satellites, suitably built up in
order to reduce the impact of the non--gravitational perturbations
by reducing, e.g., the area-to-mass ratio, with identical
eccentric orbits (say $e\sim 0.04$) in supplementary orbital
planes with $i=63.4^{\circ}$. With this choice of the inclination
the periods of many uncancelled non--gravitational perturbations
affecting \rfr{difpgi} would amount to only a few years, so that
not too long observational time spans would be required in order
to reduce the impact of such subtle perturbations. Indeed, on one
hand, if the adopted time span is a multiple of their periods they
average out and, on the other, they could be viewed as empirically
fitted quantities to be removed from the temporal series, at least
to a certain extent. In particular, it could be possible to adopt
for the data analysis just the first years of life of the
satellites: in this way many useful and simplifying assumptions on
the interplay between the satellites'spin behavior and  some
related non--gravitational perturbations of thermal origin, like
the solar Yarkovsky-Schach and the Earth infrared
Yarkovsky-Rubincam thermal thrusts could be safely done.
\subsection{Conclusions}
From quantitative investigations (Iorio, 2003a; Iorio and
Lucchesi, 2003) it turns out that such a new configuration would
yield great benefits also to the accuracy of the sum of the nodes.
Indeed, according to pessimistic evaluations, for a pair of
LAGEOS--like SLR satellites with $a=12000$ km, $e=0.04$ and
$i=63.4^{\circ}$ and a suitably chosen time span the total
systematic error in the sum of the nodes would amount to almost
$0.1\%$, which is better than that could be obtained with the
LAGEOS--LARES mission, while the total systematic error in the
difference of the perigees would be of the order of $5\%$ (without
removing any time-dependent signals and by assuming LAGEOS--like
satellites), mainly due to the non--gravitational perturbations.
It should be recalled that, when in the near future the new, more
accurate data for the Earth's gravitational field from the CHAMP
and GRACE missions will be available, the role played by the
non--gravitational perturbations in the total error budget will be
dominant. Of course, the use of drag--free satellites would be
able to greatly reduce their impact on the proposed measurement; a
data analysis time span of just a few years, as it would be
obtained with the chosen critical inclination, would be
particularly well suited in view of the finite lifetime of the
drag--free apparatus. Moreover, with the data of the orbital
elements of such new satellites it would also be possible to
follow the multiresidual linear combination approach sketched in
(Iorio $et\ al$, 2002a) in conjunction with the data from LAGEOS
and LAGEOS II.
\section{A gravitomagnetic clock effect on Earth}
\subsection{The gravitomagnetic field of the Earth}
In this section we intend to present a possible new Earth--based
laboratory experiment (Iorio, 2003c) which exploits, in a certain
sense, the concept of the gravitomagnetic clock effect (Iorio $et\
al$, 2002b) of two counter--orbiting test particles along
identical circular orbits. For other proposed gravitomagnetic
laboratory--based experiments see chapter 6 of  (Ciufolini and
Wheeler, 1995), the references in (Iorio, 2003c) and (Ruggiero and
Tartaglia, 2002).

According to the gravitational analogue of the Larmor theorem
(Mashhoon, 1993), we could obtain \rfr{eqb} by considering an
accelerated frame rotating with angular velocity \eqi
{\bm\Omega}_{\rm LT}= \frac{{\bm B}_{\rm g}}{2c}.\eqf Indeed, in
it an inertial Coriolis acceleration \eqi{\bm a}_{\rm Cor}=2{\bm
v}\times{\bm\Omega}_{\rm LT}\lb{corio}\eqf is experienced by the
proof mass. Then, in a laboratory frame attached to the rotating
Earth we could consider, apart from the true Coriolis inertial
force, also the relativistic term of \rfr{corio}. Such
gravitomagnetic feature could be measured, in principle, in the
following way.
\subsection{The experiment}
Let us choose an horizontal plane at, say, South Pole: here the
Earth's angular velocity vector ${{\bm\omega}}_{\oplus}$ and
${{\bm\Omega}}_{\rm LT}$ are perpendicular to it and have opposite
directions. Let us choose as unit vector for the $z$ axis the unit
vector $\bm{\hat{\Omega}}_{\rm LT}$, so that
\bea {{\bm\Omega}}_{\rm LT} & = & \rp{2GJ}{c^2 R_{\rm p}^3}\bm{\hat{z}},\\
{{\bm\omega}}_{\oplus} & = & -\omega_{\oplus}\bm{\hat{z}},\\
{\bm g} & = & -g\bm{\hat{z}},\eea where  ${\bm J}$ is the proper
angular momentum of the Earth, $R_{\rm p}$ is the Earth polar
radius and {\bm g} is the gravitoelectric acceleration to which,
at the poles, the centrifugal acceleration does not contribute. At
a generic latitude the same reasoning holds provided that the
component of ${{\bm\Omega}}_{\rm LT}$ along the local vertical is
considered. It is \eqi{{\bm\Omega}}_{\rm
LT}^{(v)}=-\rp{2GJ\cos\theta}{c^2 R^3}\bm{\hat{r}}\eqf where
$\theta$ is the colatitude counted from the North Pole. A particle
which moves with velocity {\bm v} in the previously considered
polar horizontal plane is acted upon by the Coriolis inertial
force induced by the noninertiality of the terrestrial reference
frame and also by the gravitational force of \rfr{corio}. Such
forces have the same line of action and opposite directions: in an
horizontal plane at South Pole the resultant acceleration is
$2v\tilde{\Omega}\equiv 2v(\Omega_{\rm LT} -\omega_{\oplus})$ and
it lies, orthogonally to ${\bm v}$, in the aforementioned plane.
Let us consider an experimental apparatus consisting of a
friction--free horizontal plane placed in a vacuum chamber. Upon
such a desk a small tungsten mass $m$, tied to a sapphire fiber of
length $l$, tension ${\bm T}$ and fixed at the other extremity, is
put in a circular uniform motion. Indeed, the forces which act on
$m$ are the tension of the wire, the Coriolis inertial force and
the Lense--Thirring gravitational force which are all directed
radially; the weight force ${\bm W}=m{\bm g}$ is balanced by the
the normal reaction ${\bm N}$ of the plane and there are neither
the atmospheric drag nor the friction of the plane. Let us assume
the counterclockwise rotation as positive direction of motion for
$m$. At the equilibrium the equation of motion is \eqi m\omega_+^2
l=T-2m\omega_+ l\tilde{\Omega},\lb{rio}\eqf where $\omega_+$ is
the angular velocity of the mass $m$ when it rotates
counterclockwise and $l$ is the radius of the circle described by
$m$. If the Earth did not rotate the angular velocity of the
particle would be \eqi\omega_0=\sqrt{\rp{T}{ml}}.\eqf  The
gravitomagnetic and the Coriolis forces slightly change such
circular frequency. Since $\omega_0>>\tilde{\Omega}$, from
\rfr{rio} it follows for both the counterclockwise and clockwise
directions of rotation
\eqi\omega_\pm=\omega_0\mp\tilde{\Omega},\eqf so that we could
adopt as observable \eqi\Delta\omega\equiv\omega_-
-\omega_+=2\tilde{\Omega}\equiv 2(\Omega_{\rm
LT}-\omega_{\oplus}).\lb{dfr}\eqf Of course, the physical
properties of the sapphire fiber and of the tungsten mass should
not change from a set of rotations in a direction to another set
of rotations in the opposite direction, so to allow an exact
cancellation of $\omega_0$ in \rfr{dfr}.
\subsection{Discussion}
Since on the Earth's surface at the poles $\Omega_{\rm
LT}=3.4\times 10^{-14}$ rad s$^{-1}$, we must ask if the
experimental sensitivity of the sketched apparatus allows to
measure such so tiny effect. If we measure the frequency shift
$\Delta\omega$ from the rotational periods of the mass $m$ we have
\eqi\delta\Omega_{\rm LT}=\rp{\delta(\Delta\omega)^{\rm
exp}+\delta\omega_{\oplus}}{2}\eqf with
\eqi\delta(\Delta\omega)^{\rm exp}=\delta\omega_-^{\rm
exp}-\delta\omega_+^{\rm exp}=2\pi\left[\left(\rp{\delta
P_-}{P^2_-}\right)^{\rm exp}-\left(\rp{\delta
P_+}{P^2_+}\right)^{\rm exp}\right].\lb{accu}\eqf

The Earth's angular velocity $\omega_{\oplus}$ is very well known
in a kinematically, dynamically independent way from the Very Long
Baseline Interferometry (VLBI) technique with an accuracy of the
order of\footnote{See on the WEB: {\bf
http://www.iers.org/iers/products/eop/long.html} and {\bf
http://hpiers.obspm.fr/eop-pc/}.} $\delta\omega_{\oplus }\sim
10^{-18}$ rad s$^{-1}$. In fact, $\omega_{\oplus}$ is not exactly
uniform and experiences rather irregular changes which are
monitored in terms of Length-Of-Day (LOD) by the Bureau
Internationale des Poids et Mesures-Time Section (BIMP) on a
continuous basis\footnote{For these topics see on the WEB: {\bf
http://einstein.gge.unb.ca/tutorial/} and {\bf
http://hpiers.obspm.fr/eop-pc/}.}. Such changes are of the order
of $\Delta\omega_{\oplus}\sim 0.25$ milliarcseconds per year (mas
yr$^{-1})=3.8\times 10^{-17}$ rad s$^{-1}$, so that they are
negligible. A possible source of error might come from our
uncertainty in the position of the proposed polar set--up with
respect to the Earth's crust, i.e. from the polar motion of the
instantaneous axis of rotation of the Earth
$\bm{\hat{\omega}}_{\oplus}$ in terms of the small angles $x$ and
$y$. It turns out that this phenomenon has three components: a
free oscillation with a (measured) period of 435 days (Chandler
Wobble) and an amplitude of less than 1 arcsecond (asec), an
annual oscillation forced by the seasonal displacement of the air
and water masses of the order of 10$^{-1}$ asecs and an irregular
drift of the order of some asecs. There are also some diurnal and
semi--diurnal tidally induced oscillations with an amplitude less
than 1 mas. As a consequence, the position of the pole is unknown
at a level of some meters. The small size of the apparatus should
overcome such problem. Moreover, it can be easily seen that the
impact of such offsets of $\bm{\hat{\omega}}_{\oplus}$ on the
Coriolis force is $2v\omega_{\oplus}\cos\delta$ with $\delta$ of
the order of some asecs or less, so that it is negligible.

In regard to the experimental measurement of the periods
$P_{\pm}$, it should be possible to strongly constrain \rfr{accu}
by choosing suitably the parameters of the apparatus so to
increase the periods and/or by measuring them after many
revolutions. However, the important point is that their difference
only is important, and it should be possible to reduce such
difference to the accuracy level required.

Of course, we are aware of the fact that many practical
difficulties would make the proposed measurement very hard to be
implemented. For example, it turns out that the the friction force
of the plane should be less than 2$\times 10^{-9}$ dyne. Moreover,
in order to reach the quoted accuracy in measuring
$\omega_{\oplus}$ with VLBI several years of continuous
observation would be required.


\begin{references}

\item{} Ciufolini I 1986 {\em Phys. Rev. Lett.} {\bf 56} 278--81\par
\item{} Ciufolini I and Wheeler J A 1995 {\em Gravitation and
Inertia} (New York: Princeton University Press)\par
\item{} Ciufolini I 2002 Test of General Relativity: 1995-2002 measurement of
frame-dragging {\em Preprint} gr-qc/0209109\par
\item{} Iorio L 2003a On a new observable for measuring the Lense-Thirring
effect with Satellite Laser Ranging  {\em Gen. Rel. and Gravit.}
{\bf 35} 1583--95\par
\item{}\dash 2003b A new proposal for measuring the
Lense-Thirring effect with a pair of supplementary satellites in
the gravitational field of the Earth  {\em Phys. Lett.} A {\bf 308
} 81--4\par
\item{}\dash 2003c On the possibility of measuring the Earth's gravitomagnetic
force in a new laboratory experiment {\em Class. Quantum Grav.}
{\bf 20} L5--9\par
\item{} \dash 2003d The impact of the static part of the Earth's gravity field
on some tests of General Relativity with Satellite Laser Ranging
{\em Celest. Mech.} {\bf 86} 277--94\par
\item{} Iorio L and Lucchesi D M 2003 LAGEOS-type Satellites in Critical
Supplementary Orbital Configuration and the Lense-Thirring Effect
Detection {\em Class. Quantum Grav.} {\bf 20} 2477--90\par
\item{} Iorio L, H I M Lichtenegger and Mashhoon B 2002 An alternative derivation of the
gravitomagnetic clock effect, {\em Class. Quantum Grav.} {\bf 19}
39--49\par
\item{} Iorio L, Lucchesi D M and Ciufolini I 2002 The LARES
mission revisited: an alternative scenario, {\em Class.  Quantum
Grav.} {\bf 19} 4311-25\par
\item{} Kaula W M 1966 {\em Theory of Satellite
Geodesy} (Waltham: Blaisdell Publishing Company)\par
\item{} Mashhoon B 1993 On the Gravitational
Analogue of Larmor Theorem {\em Phys. Lett.} A {\bf 173}
347--54\par
\item{} Ruggiero M L and Tartaglia A 2002 Gravitomagnetic effects
{\em Nuovo Cimento} B {\bf 117} 743--67\par



\end{references}
\end{document}